\documentclass[aps,prl,twocolumn,superscriptaddress,showpacs,longbibliography]{revtex4-2}

\usepackage{graphicx}
\usepackage{amsmath}
\usepackage{amssymb}
\usepackage{bm}
\usepackage{physics}
\usepackage{mathtools}
\usepackage{xcolor}
\usepackage{hyperref}
\usepackage{lineno}

\usepackage[final]{changes}

\begin{document}



\title{Localization in microcavities revealed by phase-space non-Hermitian skin effect}


\author{Jung-Wan Ryu}
    \affiliation{Center for Trapped Ion Quantum Science, Institute for Basic Science (IBS), Daejeon 34126, Republic of Korea}
    \affiliation{Basic Science Program, Korea University of Science and Technology (UST), Daejeon 34113, Republic of Korea}

\author{Yong-Hoon Lee}
    \affiliation{School of Electronic and Electrical Engineering, Kyungpook National University, Daegu 41566, Republic of Korea}

\author{Muhan Choi}
    \affiliation{School of Electronic and Electrical Engineering, Kyungpook National University, Daegu 41566, Republic of Korea}
    \affiliation{Digital Technology Research Center, Kyungpook National University, Daegu 41566, Republic of Korea}

\author{Chang-Hwan Yi}
    \email{yichanghwan@hanmail.net}
    \affiliation{Center for Theoretical Physics of Complex Systems, Institute for Basic Science (IBS), Daejeon 34126, Republic of Korea}
    \affiliation{Department of Physics, Pukyong National University, Busan 48513, Republic of Korea}
    \affiliation{Department of Physics, Hanyang University, Seoul 04763, Republic of Korea}

\author{Martina Hentschel}
    \email{martina.hentschel@physik.tu-chemnitz.de}
    \affiliation{Institute of Physics, Technische Universit{\"a}t Chemnitz, D-09107 Chemnitz, Germany}


\begin{abstract}
Contrary to the semiclassical expectation for fully chaotic systems, localization of resonances is found to be a common feature in open microcavities. In spiral-shaped dielectric microcavities, a substantial fraction of resonances localize on polygonal patterns in real space, are chiral, and their momentum distributions accumulate near the critical line for total internal reflection. Despite the extensive investigation, the physical mechanism responsible for their remarkable abundance has remained a long-standing question. Addressing this, we reveal a physical correspondence between an inhomogeneous-loss Hatano-Nelson model and the dielectric phase space of a spiral microcavity. We show that the combination of geometry-induced momentum drift and refractive escape yields a generalized non-Hermitian skin effect in the phase space momentum. We identify this mechanism as the origin of the critical-line localization of resonances in open chaotic spiral microcavities, extending the skin-effect concept beyond nonreciprocal lattices to phase space and to open chaotic wave systems.
\end{abstract}


\maketitle


Localization is \replaced{among}{one of} the most fundamental wave phenomena and occurs in a wide variety of systems \cite{Anderson1958absence, Evers2008anderson, Segev2013anderson, Lagendijk2009fifty}. Though it is a universal feature, the details of its manifestation depend on system properties such as the external potential, the internal dynamics, or the degree of openness. In open chaotic systems including the well-studied deformed microcavities, openness fundamentally modifies the underlying wave dynamics and can give rise to localization behavior qualitatively different from that of the corresponding closed systems \cite{Noeckel1997ray, Gmachl1998high-power, Cao2015dielectric}. 

\begin{figure}[t]
\centering
\includegraphics[width=\linewidth]{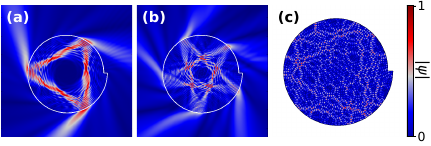}
\caption{
Representative localized resonances in spiral dielectric microcavities and the corresponding closed billiard.
(a) A triangularly localized resonance for $n=2$.
(b) A star-shaped localized resonance for $n=3$.
(c) A representative weakly localized eigenstate in the corresponding closed billiard for $n=2$.
Strongly localized resonances constitute a significant fraction of the resonance spectrum in spiral dielectric microcavities, whereas comparable localization is rare in the corresponding closed billiard.
}
\label{fig:localized}
\end{figure}

One of \added{the} most striking examples demonstrating how openness-induced changes in the wave dynamics manifest themselves in qualitatively different localization properties is the spiral dielectric microcavity. It supports an unusually large number of strongly localized resonances [see Figs.~\ref{fig:localized}(a) and (b)] with momentum distributions that systematically accumulate near the critical line for total internal reflection (TIR) \cite{Lee2004quasiscarred}. 
\replaced{The physical mechanism responsible for the abundance of resonances predominantly localized near the critical line has remained unresolved.}{The physical mechanism responsible for the abundance of localized resonances and their systematic accumulation near the critical line has remained unresolved.}

Since the first observation of so-called quasi-scarred resonances in spiral dielectric microcavities, numerous studies have revealed a variety of unusual wave phenomena in these systems. The characteristic polygonal wave patterns of such strongly localized resonances have been attributed to wave corrections \replaced{of}{to} the underlying ray dynamics \cite{Goos1947Ein, Tureci2002deviation, Schomerus2006correcting, Unterhinninghofen2008Goos-Hanchen, Altmann2008non-hamiltonian, Unterhinninghofen2010interplay} and to coupling and mixing between resonance modes \cite{Wiersig2006formation, Liu2006wave, Lee2008resonances, Yang2009optical, Yi2015fermi}. Their preferred propagation direction has been related to the asymmetric ray dynamics generated by the spiral geometry \cite{Lee2025scarred, Schafer2006directed, Lee2008ray}. The broken chiral symmetry of spiral cavities has also been connected to the emergence of nonorthogonal resonance pairs \cite{Wiersig2008asymmetric, Kullig2016frobenius-perron}. Classical steady-survival distributions and conditionally invariant measures have shown that the average structure and lifetime dependence of resonance modes in open chaotic wave systems can be understood from the underlying classical phase-space dynamics \cite{Harayama2015ray-wave, Clauss2019structure, Ketzmerick2022chaotic, Ketzmerick2025semiclassical}. These studies, however, do not provide a common mechanism underlying the abundance of localized resonances and their systematic accumulation near the critical line, and the case in which no suitable polygonal resonance pattern exists for a given refractive index has not been addressed.

Figure~\ref{fig:localized} compares representative states in the spiral dielectric microcavity and the corresponding closed billiard. While the dielectric microcavity supports strongly localized polygonal resonances [Figs.~\ref{fig:localized}(a) and (b)], the corresponding closed billiard exhibits only weak localization [Fig.~\ref{fig:localized}(c)], that is moreover increasingly difficult to identify as the semiclassical parameter increases \cite{Lee2004quasiscarred, Lee2008ray, Rim2018quantum}. Additional resonance patterns presented in Sec.~S1 of the Supplemental Material further demonstrate that such localized resonances constitute a substantial fraction of the resonance spectrum. This striking contrast indicates that dielectric openness plays an essential role in the localization mechanism.

The Hatano-Nelson model \cite{Hatano1996localization, Hatano1997vortex} provides a simple setting in which asymmetric hopping and an open boundary give rise to the non-Hermitian skin effect (NHSE), resulting in the macroscopic accumulation of eigenstates near a boundary \cite{Kunst2018biorthogonal, Yao2018edge, Lee2019anatomy, Yokomizo2019non-bloch, Okuma2020topological, Zhang2020correspondence, Helbig2020generalized, Xiao2020non-hermitian, Weidemann2020topological, Bergholtz2021exceptional, Zhang2022areview, Wang2023experimental, Hu2024geometric, Li2024observation, Yoshida2024non-hermitian, Zhao2025two-dimensional}. This raises the possibility that the long-standing localization problem in spiral dielectric microcavities could share the same underlying mechanism as the NHSE. Here we establish a physical correspondence between an inhomogeneous-loss Hatano-Nelson model and the dielectric phase space of a spiral microcavity. We show that the spiral geometry generates momentum drift in dielectric phase space, while refractive escape creates an effective boundary in phase space at the critical line for TIR. Neither ingredient alone is sufficient to produce interface localization; their combination gives rise to a generalized NHSE in dielectric phase space. This mechanism explains both the remarkable abundance of localized resonances and their systematic accumulation near the critical line. Our work extends the concept of the NHSE beyond nonreciprocal lattice systems and beyond real space and identifies a general physical mechanism for localization in open chaotic wave systems with geometry-induced momentum drift and loss-induced effective boundaries.

\begin{figure}[t]
\centering
\includegraphics[width=\linewidth]{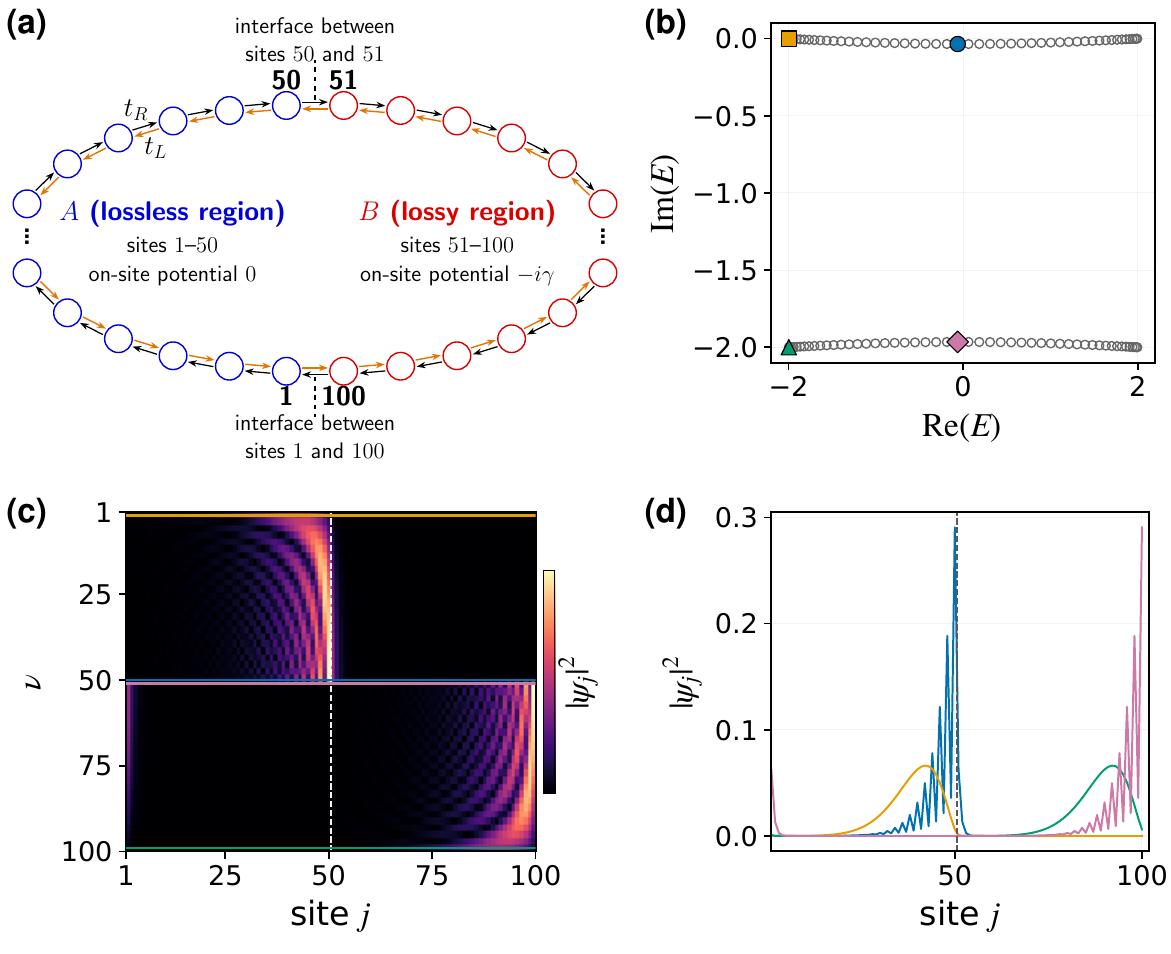}
\caption{
Hatano-Nelson model with spatially inhomogeneous loss.
(a) Schematic illustration of a periodic ring consisting of a lossless region $A$ and a lossy region $B$.
(b) Complex-energy spectrum.
(c) Spatial intensity distributions of all right eigenstates, indexed by \(\nu\) in order of decreasing $\mathrm{Im}(E)$.
(d) Representative right eigenstates selected from the weakly and strongly decaying branches.
Despite the periodic boundary conditions, the right eigenstates exhibit pronounced interface localization near the two interfaces separating the lossless and lossy regions.
}
\label{fig:HN_numerics}
\end{figure}

\textit{\textcolor{blue}{Inhomogeneous-loss Hatano-Nelson model.}}
To isolate the minimal ingredients responsible for the localization, we first consider a periodic Hatano-Nelson ring with spatially inhomogeneous loss.
As shown in Fig.~\ref{fig:HN_numerics}(a), the ring consists of a lossless region $A$ with $N_A$ sites and a lossy region $B$ with $N_B$ sites, where $N=N_A+N_B$.
Its Hamiltonian is
\begin{equation}
H=
\sum_{j=1}^{N}
\left(
t_R |j+1\rangle\langle j|
+
t_L |j\rangle\langle j+1|
\right)
-
i\gamma
\sum_{j=N_A+1}^{N}
|j\rangle\langle j|,
\label{eq:HN}
\end{equation}
where the site index is understood modulo $N$, the asymmetric hopping amplitudes are
$t_R=t e^{g}$,
$t_L=t e^{-g}$,
and $g>0$ biasing the hopping toward increasing site indices.
Unlike the conventional Hatano-Nelson model, the present system combines asymmetric hopping with spatially inhomogeneous loss while preserving periodic boundary conditions.
Throughout this work we use
$N=100$,
$N_A=N_B=50$,
$g=0.1$,
and
$\gamma=2$.

Figures~\ref{fig:HN_numerics}(b)-\ref{fig:HN_numerics}(d) show the resulting complex spectrum and right eigenstates.
The complex-energy spectrum separates into two well-defined branches, with the strongly decaying branch shifted downward by approximately $-i\gamma$ \replaced{due}{owing} to the onsite loss.
Figure~\ref{fig:HN_numerics}(c) demonstrates that this spectral separation is accompanied by a simultaneous separation in the spatial structure of the eigenstates: every right eigenstate is localized near one of the two interfaces despite the periodic boundary conditions.
Within each branch, the eigenstates share a common exponentially localized envelope, while their nodal structures depend on the mode index, as illustrated in Fig.~\ref{fig:HN_numerics}(d).


\textit{\textcolor{blue}{Origin of interface localization.}}
The physical origin of the interface localization becomes transparent in the strong-loss limit $\gamma\gg t$.
In this limit, hybridization between the lossless and lossy regions is strongly suppressed, leading to a clear separation into weakly and strongly decaying branches supported predominantly in the lossless and lossy regions, respectively.
Consequently, each branch is described by an effective open-boundary Hatano-Nelson chain defined on the corresponding region.
The standard open-boundary Hatano-Nelson solution then gives the $m$-th right eigenstate of the inhomogeneous-loss model in the strong-loss limit as
\begin{equation}
\psi_m(\ell)
=
e^{g\ell}
\sin\!\left(
\frac{m\pi \ell}{N_\alpha+1}
\right),
\end{equation}
where \(\ell=1,\ldots,N_\alpha\) is the site index within the corresponding region, with \(\alpha=A\) for the weakly decaying branch and \(\alpha=B\) for the strongly decaying branch.
The sinusoidal factor determines the mode-dependent nodal structure, whereas the exponential envelope represents the NHSE.
The detailed derivation, together with the finite-$\gamma$ corrections, is presented in End Matter A.

Starting from weak loss, increasing the spatially inhomogeneous loss produces a width bifurcation into weakly and strongly decaying branches.
The weakly decaying branch acquires progressively smaller decay widths while becoming increasingly confined to the lossless region, corresponding to resonance trapping induced by spatially inhomogeneous loss.
Asymmetric hopping then localizes these states toward the interface through the NHSE.


\textit{\textcolor{blue}{Dielectric phase-space realization.}}
We now show that the dielectric phase space of a spiral microcavity realizes the same localization mechanism without an underlying lattice or explicit asymmetric hopping.


\begin{figure}[t]
\centering
\includegraphics[width=\linewidth]{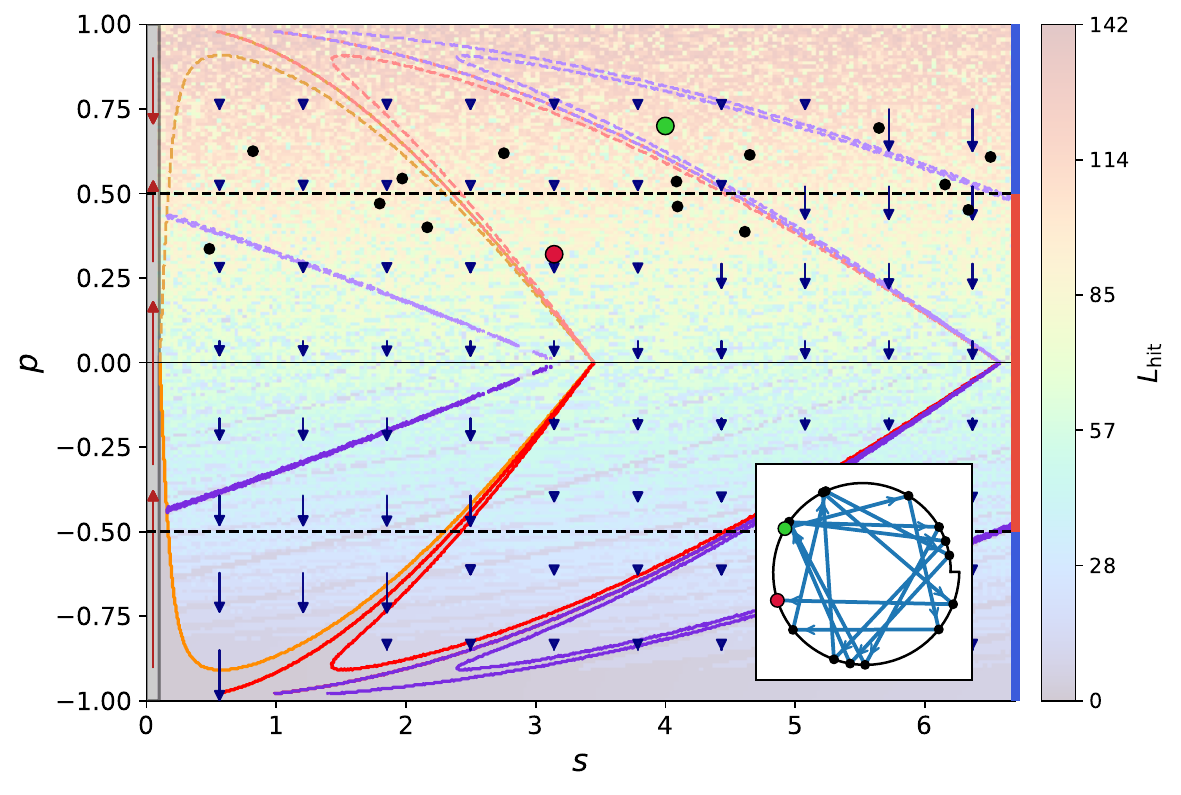}
\caption{Poincar\'e surface of section of the spiral billiard.
The horizontal coordinate \(s\) is the arclength coordinate along the entire cavity boundary, \replaced{starting from the notch and then following the spiral arc}{including both the spiral arc and the notch}, and \(p=\sin\chi\) is the Birkhoff momentum coordinate.
The background color represents \(L_{\rm hit}\), \replaced{length of trajectory started at}{the trajectory length from} the spiral arc to the first encounter with the notch, and the gray-shaded region denotes the notch.
Blue and red segments on the right boundary separate the total-internal-reflection region, \(|p|>p_c\), and the refractive escape region, \(|p|<p_c\), respectively.
Blue arrows indicate the momentum change, \(\Delta p=p_{n+1}-p_n\), between successive \replaced{reflections}{collisions} on the spiral arc, whereas red arrows indicate the corresponding momentum change in the notch region.
Solid curves indicate phase-space regions whose trajectories reach the notch after one, two, and three subsequent reflections, whereas dashed curves indicate regions reached after one, two, and three reflections following emergence from the notch.
The inset shows a representative ray trajectory and the corresponding collision sequence in phase space.
}
\label{fig:spiral_flow}
\end{figure}

Figure~\ref{fig:spiral_flow} shows the Poincar\'e surface of section of the spiral billiard, where \(s\) measures arclength along the entire cavity boundary, corresponding to $0 < s < \epsilon$ for the notch and $\epsilon < s <s_{max}$ for the spiral arc, where $\epsilon$ is the length of the notch, and \(p=\sin\chi\) is the Birkhoff momentum coordinate \cite{Stockmann1999quantum, Tabachnikov2005geometry}.
The smooth spiral arc and the notch play qualitatively different roles in the phase-space dynamics.
To isolate the transport generated by the spiral arc, we consider the arc-to-arc component of the one-step billiard map, for which both the present and subsequent collisions occur on the smooth spiral boundary.
Each blue arrow represents the change in momentum between two successive collisions on the spiral arc, starting from a phase-space point $(s_n,p_n)$ and indicating the direction and relative magnitude of $\Delta p=p_{n+1}-p_n$. The systematic downward orientation of the blue arrows therefore directly visualizes the drift toward smaller $p$.
This drift is a deterministic consequence of the distinctive spiral geometry.
Because the radius and the local boundary orientation vary continuously along the spiral arc, successive reflections modify the incidence angle asymmetrically, biasing the billiard map toward decreasing $p$.
The inset illustrates the corresponding collision sequence of a representative ray in real space, while the background color shows the trajectory length to the first notch encounter, providing a complementary global view of the resulting phase-space transport.

The notch interrupts the systematic drift and reinjects trajectories into the spiral arc. The solid curves separate regions whose trajectories reach the notch after one, two, and three subsequent reflections, while the dashed curves show their images after emergence from the notch. The resulting cycle of arc-induced momentum drift and notch reinjection forms a global circulation in the closed spiral billiard. This circulation is analogous to the directionally biased propagation around the periodic Hatano-Nelson ring before spatially inhomogeneous loss is introduced. Since the circulation remains uninterrupted in the closed billiard, the geometry-induced momentum drift alone does not produce systematic localization.

Dielectric openness fundamentally changes this circulation.
For a cavity with refractive index \(n\), the critical momentum is
\begin{equation}
p_c=\frac{1}{n},
\label{eq:critical_momentum}
\end{equation}
and TIR occurs for \(|p|>p_c\).
The region \(|p|>p_c\) supports TIR, whereas trajectories entering \(|p|<p_c\) lose part of their intensity through refractive escape governed by the Fresnel reflection coefficient \cite{Born1999principles}.
The blue and red segments on the right boundary of Fig.~\ref{fig:spiral_flow} visualize the total-internal-reflection (trapped) and refractive regions and directly parallel the lossless and lossy regions in Fig.~\ref{fig:HN_numerics}(a).
Once the geometry-induced drift carries a wave component across the positive critical line, refractive escape attenuates it before the closed phase-space circulation can be completed. The critical line therefore acts as an effective open boundary for the long-lived wave components.

\textit{\textcolor{blue}{From directed wave propagation to asymmetric hopping.}}
\added{To complete the correspondence with the Hatano-Nelson model, we now show how the geometry-induced momentum drift gives rise to effective asymmetric hopping at the wave level. We account for the finite spatial extent associated with wave propagation by replacing the ray-propagation kernel with a localized kernel of width \(w\) and projecting it onto localized basis states of width parameter \(\sigma\),
separated by a distance \(d\). For a propagation kernel whose center is displaced by \(a>0\) along the
transport direction, the resulting forward and backward hopping amplitudes, \(t_R\) and \(t_L\), respectively, satisfy
\begin{equation}
\frac{t_R}{t_L}
=
\exp\left(
\frac{2ad}{\Sigma^2}
\right)
>1,
\label{eq:hopping_ratio}
\end{equation}
where \(\Sigma^2=w^2+2\sigma^2\), with \(\Sigma\) representing the effective width resulting from the propagation kernel and the localized basis states. Thus, the geometry-induced momentum drift translates at the wave level into an effective Hatano-Nelson description with asymmetric hopping. The detailed derivation is presented in End Matter B and Supplemental Material~S2.}

\begin{table}[t]
\centering
\resizebox{\columnwidth}{!}{%
\begin{tabular}{c|c}
\hline
Hatano-Nelson model &
Spiral dielectric phase space\\
\hline
asymmetric hopping &
geometry-induced momentum drift\\

periodic lattice coordinate &
Birkhoff momentum coordinate \(p\)\\

lossless region &
\(|p|>p_c\)\\

lossy region &
\(|p|<p_c\)\\

effective open boundary &
critical line \(|p|=p_c\)\\

spatially inhomogeneous loss &
dielectric escape\\
\hline
\end{tabular}%
}

\caption{
Correspondence between the inhomogeneous-loss Hatano-Nelson model and the dielectric phase space of a spiral microcavity.
}
\label{tab:mapping}
\end{table}

The correspondence summarized in Table~\ref{tab:mapping} identifies the accumulation of long-lived wave components on the trapped side of the critical line as the phase-space counterpart of interface localization in the inhomogeneous-loss Hatano-Nelson model.
In the lattice model, asymmetric hopping drives wave components toward an interface separating regions with different losses, whereas in the spiral cavity geometry-induced drift drives them toward the critical line separating the trapped and refractive regions.
Without dielectric escape, the geometry-induced drift merely generates the global circulation of the closed billiard rather than systematic localization.
The combination of geometry-induced drift and dielectric escape therefore produces systematic accumulation of long-lived wave components near the critical momentum $p_c = 1/n$.

The resulting phenomenon is a generalized NHSE.
Unlike the conventional skin effect, where asymmetric hopping localizes eigenstates along a real-space coordinate, the accumulation here occurs instead along the momentum coordinate of dielectric phase space.
Moreover, the effective open boundary originates from refractive escape rather than from a physical termination of the system.
This interpretation distinguishes the general localization mechanism from the formation of specific polygonal wave patterns.
The generalized skin effect explains the abundance of long-lived resonances localized near the critical line, whereas triangular, star-shaped, and other polygonal resonances require additional geometrical commensurability conditions.
The polygonal resonances are therefore specific manifestations of the more general critical-line localization mechanism.


\textit{\textcolor{blue}{Statistical verification.}}
The generalized skin-effect picture predicts that localization near the critical line is a common property of the resonance spectrum rather than a feature of a few isolated modes.
To examine this prediction, we consider refractive indices \(n=2\), \(3\), and \(4\) and analyse 1000 resonances with $nkR \sim 100$ for each case.
In the inhomogeneous-loss Hatano-Nelson model, sufficiently strong loss separates the spectrum into weakly and strongly decaying branches through width bifurcation.
The corresponding eigenstates become predominantly supported in the lossless and lossy regions, respectively, and asymmetric hopping subsequently localizes each branch toward the corresponding interface.
In dielectric microcavities, however, the effective dielectric loss is not sufficiently strong to produce a clearly resolved separation of the two branches.
We therefore identify the predominantly trapped resonances statistically rather than by separating distinct spectral branches.

For each resonance, the boundary Husimi distribution \(H(s,p)\) is averaged over the boundary coordinate \cite{Husimi1940some, Leboeuf1990chaos-revealing, Crespi1993quantum, Hentschel2003husimi}, and the momentum at which the averaged Husimi distribution attains its maximum is denoted by
\begin{equation}
\overline{H}(p)
=
\frac{1}{L_{\partial}}
\int_0^{L_{\partial}}
H(s,p)\,ds,
\qquad
p_{\max}
=
\underset{p}{\operatorname{arg\,max}}\,
\overline{H}(p),
\label{eq:pmax}
\end{equation}
where \(L_{\partial}\) is the total boundary length. The quantity \(p_{\max}\) serves as a measure of the position of the dominant localization along the phase-space momentum coordinate. Characterizing a resonance by its averaged momentum can obscure the dominant localization, either because resonances localized near the critical line generally exhibit asymmetric localization profiles or because multiple peaks contribute simultaneously (Supplemental Material~S3). Since most resonances localized near the critical line exhibit a single dominant Husimi peak, \(p_{\max}\) provides a direct and robust characterization of the dominant localization.

\begin{figure}[t]
\centering \includegraphics[width=\linewidth]{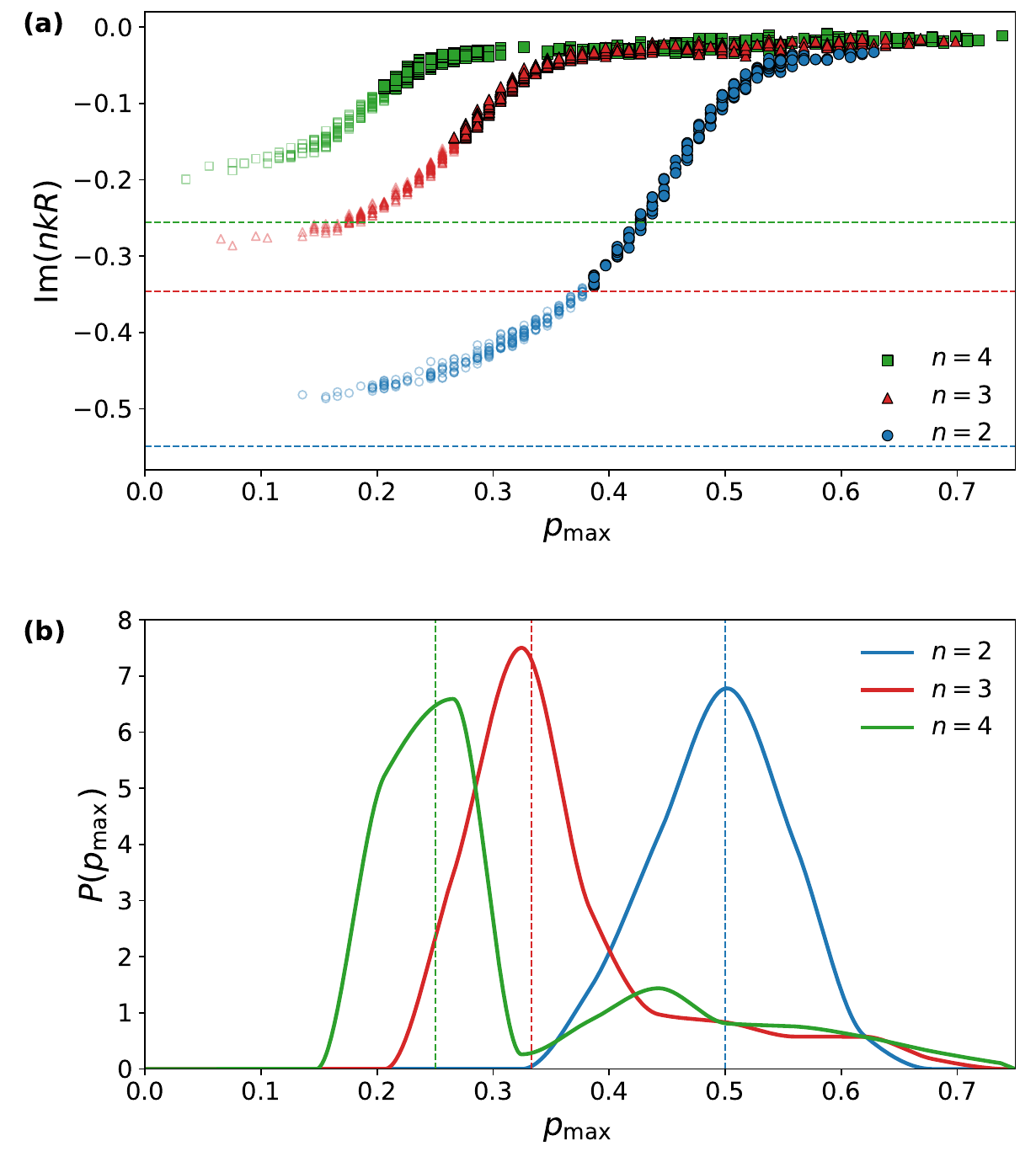}
\caption{Statistics of the localization properties of \(1000\) resonances of the spiral dielectric microcavity for each refractive index \(n=2\), \(3\), and \(4\).
(a) Imaginary part of the resonance wavenumber, $\mathrm{Im}(nkR)$, as a function of \(p_{\max}\), the momentum at which the boundary Husimi distribution reaches its maximum.
Dark symbols denote the long-lived resonances selected for the statistical analysis, whereas light symbols represent the remaining internal resonances above the lower bound on $\mathrm{Im}(nkR)$.
Horizontal dashed lines indicate the corresponding lower bounds, $\operatorname{Im}(nkR)=\ln\left|({n-1})/({n+1})\right|/2$.
(b) Probability density \(P(p_{\max})\) of \(p_{\max}\) for the selected long-lived resonances.
Vertical dashed lines denote the critical momenta, $p_c=1/n$.}
\label{fig:pmax_distribution}
\end{figure}

Figure~\ref{fig:pmax_distribution}(a) shows \(\mathrm{Im}(nkR)\) as a function of \(p_{\max}\).
The dark symbols denote the relatively long-lived resonances selected for the statistical analysis, while the light symbols represent the remaining internal resonances.
The selection procedure, including the lower bound on \(\mathrm{Im}(nkR)\) used to identify internal resonances and the phase-space estimate of the number of total-internal-reflection modes, is described in End Matter C.
The probability densities in Fig.~\ref{fig:pmax_distribution}(b) show that a large fraction of the selected resonances accumulates near the positive critical momentum for every refractive index considered.
This behavior is not restricted to n=2 and n=3, where triangular and star-shaped resonances are prominent, but also occurs for n=4, showing that critical-line localization is not tied to a well-defined polygonal resonance family. These statistical results confirm the phase-space mechanism illustrated in Fig.~\ref{fig:spiral_flow} and show that the abundance of critical-line-localized resonances is a robust consequence of the generalized NHSE in dielectric phase space.

\added{To understand why some selected resonances exhibit localization maxima inside the refractive region, \(p_{\max}<p_c\), unlike the weakly decaying states of the sharp-interface Hatano-Nelson model
[cf. Fig.~\ref{fig:HN_numerics}], we examine the effect of the dielectric loss profile on the Hatano-Nelson model. In an actual dielectric cavity, the escape rate varies continuously with the incidence angle according to the Fresnel law rather than changing discontinuously at the critical line. Incorporating the corresponding momentum-dependent loss into the Hatano-Nelson model reproduces these localization maxima, particularly for the relatively shorter-lived states, while leaving the underlying generalized non-Hermitian skin-effect mechanism unchanged (Supplemental Material~S4).}


In summary, we have identified the physical mechanism underlying the abundance of localized resonances in spiral dielectric microcavities. An inhomogeneous-loss Hatano-Nelson model shows that asymmetric hopping and spatially inhomogeneous loss produce localization at the interface between lossless and lossy regions. The dielectric phase space realizes the same mechanism: the spiral geometry generates a systematic momentum drift, while refractive escape turns the critical line into an effective open boundary. The corresponding closed billiard retains the drift but exhibits only weak and rare localization, showing that the localization observed in the open dielectric system requires the interplay of geometry-induced momentum drift and dielectric openness. The resonance statistics further show that this localization persists across refractive indices with and without prominent polygonal resonance families. These results establish critical-line localization in spiral microcavities as a generalized non-Hermitian skin effect in dielectric phase space, extending the skin-effect concept beyond nonreciprocal lattices.


\begin{acknowledgments}
We acknowledge financial support from the Institute for Basic Science in the Republic of Korea through projects IBS-R041-A2-2026-a00 and IBS-R041-D1-2026-a00. C.-H. Yi acknowledges financial support from the
National Research Foundation of Korea (NRF), funded by the Korean government (MSIT), under Grants No. RS-2025-16070482, RS-2025-25464760, RS-2023-NR119928, RS-2025-25446099, RS-2025-03392969, RS-2023-00278511, and RS-2025-02315685.
\end{acknowledgments}

\bibliography{references}



\onecolumngrid
~
\begin{center}
\textbf{\large End Matter}
\end{center}

\twocolumngrid

\appendix

\section{A. Analytical derivation of the inhomogeneous-loss Hatano-Nelson model}
\label{app:HN}

This End Matter provides the analytical derivation underlying the inhomogeneous-loss Hatano-Nelson model discussed in the main text.
We first recall the right eigenstates of the open-boundary Hatano-Nelson chain and then derive the effective Hamiltonians of the inhomogeneous-loss model in the strong-loss limit.


\subsection{Open-boundary Hatano-Nelson chain}

Applying the similarity transformation
\begin{equation}
\psi_j=e^{gj}\phi_j,
\end{equation}
maps the open-boundary Hatano-Nelson chain to a Hermitian nearest-neighbor tight-binding chain.
For a chain of length \(L\), the resulting right eigenstates are
\begin{equation}
\psi_m(j)
=
e^{gj}
\sin
\left(
\frac{m\pi j}{L+1}
\right),
\label{eq:appendix_HN_exact}
\end{equation}
which exhibit the characteristic exponential envelope of the NHSE.


\subsection{Strong-loss limit}

The Hamiltonian of the inhomogeneous-loss model can be written as
\begin{equation}
H=
\begin{pmatrix}
H_A & V_{AB}\\
V_{BA} & H_B-i\gamma I
\end{pmatrix},
\end{equation}
where regions \(A\) and \(B\) denote the lossless and lossy parts of the lattice, respectively.
For the weakly decaying branch,
\begin{equation}
\psi_B
=
O(\gamma^{-1})\psi_A,
\end{equation}
which gives the effective Hamiltonian
\begin{equation}
H_A^{\rm eff}
=
H_A
-
\frac{i}{\gamma}
V_{AB}V_{BA}
+
O(\gamma^{-2}).
\end{equation}
Thus,
\begin{equation}
H_A^{\rm eff}
\rightarrow
H_A,
\qquad
(\gamma\rightarrow\infty),
\end{equation}
corresponding to an open-boundary Hatano-Nelson chain on the lossless region.

Similarly, for the strongly decaying branch,
\begin{equation}
\psi_A
=
O(\gamma^{-1})\psi_B,
\end{equation}
leading to
\begin{equation}
H_B^{\rm eff}
=
H_B
+
\frac{i}{\gamma}
V_{BA}V_{AB}
+
O(\gamma^{-2}).
\end{equation}
After removing the uniform shift \(-i\gamma\), the dynamics again reduces to an open-boundary Hatano-Nelson chain on the lossy region.

Therefore, in the strong-loss limit, the original periodic inhomogeneous-loss lattice separates into two effective open-boundary Hatano-Nelson chains of lengths \(N_A\) and \(N_B\), corresponding to the weakly and strongly decaying branches, respectively.
Applying Eq.~(\ref{eq:appendix_HN_exact}) to each effective chain, with \(L\) replaced by \(N_\alpha\) (\(\alpha=A,B\)), yields Eq.~(2) of the main text.
Since the finite-\(\gamma\) corrections are confined to the interfaces, the localization mechanism remains robust provided that the two spectral branches remain well separated.

\section{B. Effective Hatano-Nelson description from directed wave propagation}
\label{app:wave_hopping}

Here we provide the derivation underlying the correspondence between the
geometry-induced momentum drift and the asymmetric hopping of the effective
Hatano-Nelson description.
The purpose of this construction is not to provide a microscopic description
of wave propagation in the spiral cavity, but to show how directed propagation,
when projected onto localized basis states with finite spatial overlap,
generates asymmetric hopping.

Consider the directed map
\begin{equation}
x_{n+1}=x_n+a,
\qquad
a>0,
\end{equation}
whose classical propagation is represented by
\begin{equation}
K_{\rm cl}(x,x')
=
\delta(x-x'-a).
\end{equation}
To account for the finite spatial extent of wave propagation, we replace the
delta-function kernel by a spatially localized kernel of finite width,
\begin{equation}
K(x,x')
=
G(x-x'-a),
\end{equation}
where \(G(y)\) is taken to be a Gaussian with width parameter \(w\),
\begin{equation}
G(y)
=
\frac{1}{\sqrt{2\pi}w}
\exp
\left(
-\frac{y^2}{2w^2}
\right).
\end{equation}

We introduce localized basis states centered at \(x_j=jd\),
\begin{equation}
\phi_j(x)
=
\frac{1}{(\pi\sigma^2)^{1/4}}
\exp
\left[
-\frac{(x-x_j)^2}{2\sigma^2}
\right],
\qquad
x_j=jd,
\end{equation}
where \(d\) is the spacing between neighboring basis states and
\(\sigma\) is their width parameter.
The projected matrix elements of the one-step propagation operator are
\begin{equation}
t_{ij}
=
\langle
\phi_i
|
K
|
\phi_j
\rangle.
\end{equation}
Straightforward Gaussian integration yields
\begin{equation}
t_{ij}
=
t_0
\exp
\left[
-\frac{(x_i-x_j-a)^2}
{2\Sigma^2}
\right],
\end{equation}
where
\begin{equation}
\Sigma^2=w^2+2\sigma^2,
\qquad
t_0=\frac{\sqrt{2}\sigma}{\Sigma}.
\end{equation}

Retaining only nearest-neighbor matrix elements gives
\begin{equation}
t_R
=
t_0
\exp
\left[
-\frac{(d-a)^2}
{2\Sigma^2}
\right],
\end{equation}
and
\begin{equation}
t_L
=
t_0
\exp
\left[
-\frac{(d+a)^2}
{2\Sigma^2}
\right].
\end{equation}
Their ratio is therefore
\begin{equation}
\frac{t_R}{t_L}
=
\exp
\left(
\frac{2ad}{\Sigma^2}
\right)
>1,
\end{equation}
which is the result used in the main text.

The asymmetric nearest-neighbor matrix elements can be parameterized in the
standard Hatano-Nelson form,
\begin{equation}
t_R
=
t e^{g_{\rm eff}},
\qquad
t_L
=
t e^{-g_{\rm eff}},
\end{equation}
with
\begin{equation}
t
=
t_0
\exp
\left[
-\frac{d^2+a^2}
{2\Sigma^2}
\right],
\qquad
g_{\rm eff}
=
\frac{ad}{\Sigma^2}.
\end{equation}
Thus, directed propagation projected onto localized basis states with finite
spatial overlap generates asymmetric hopping.
Under the localized-basis and nearest-neighbor approximations, the resulting
wave propagation is described by an effective Hatano-Nelson model.
The detailed evaluation of the Gaussian projection is presented in
Supplemental Material~S2.

\section{C. Selection of long-lived resonances}

The statistical analysis in Fig.~4 is restricted to long-lived resonances.
Since the weakly and strongly decaying branches are not clearly separated in spiral dielectric microcavities, the long-lived subset is selected statistically.

Resonances satisfying
\begin{equation}
\operatorname{Im}(nkR)
<
\frac12
\ln\left|
\frac{n-1}{n+1}
\right|
\end{equation}
are first excluded because they lie below the normal-incidence Fresnel limit.
The remaining resonances are ordered according to \(\operatorname{Im}(nkR)\).

Assuming an approximately uniform momentum density, the phase-space fraction corresponding to TIR is
\begin{equation}
f_{\rm TIR}=1-\frac1n.
\end{equation}
Accordingly, the upper fraction \(1-1/n\) of the remaining resonances is used for the statistical analysis in Fig.~4.


\clearpage
\onecolumngrid

\makeatletter
\@removefromreset{equation}{section}
\let\thesubsection\SMnormal@thesubsection
\let\thesubsubsection\SMnormal@thesubsubsection
\makeatother

\setcounter{section}{0}
\setcounter{subsection}{0}
\setcounter{subsubsection}{0}
\setcounter{figure}{0}
\setcounter{table}{0}
\setcounter{equation}{0}
\setcounter{footnote}{0}

\setcounter{secnumdepth}{3}
\renewcommand{\thesection}{S\arabic{section}}
\renewcommand{\thefigure}{S\arabic{figure}}
\renewcommand{\thetable}{S\arabic{table}}
\renewcommand{\theequation}{S\arabic{equation}}

\makeatletter
\@ifpackageloaded{hyperref}{%
  \renewcommand{\theHsection}{SM.\arabic{section}}
  \renewcommand{\theHsubsection}{SM.\arabic{section}.\arabic{subsection}}
  \renewcommand{\theHsubsubsection}{SM.\arabic{section}.\arabic{subsection}.\arabic{subsubsection}}
  \renewcommand{\theHfigure}{SM.\arabic{figure}}
  \renewcommand{\theHtable}{SM.\arabic{table}}
  \renewcommand{\theHequation}{SM.\arabic{equation}}
}{}
\makeatother

\begingroup
\centering
\vspace*{0.5em}
{\large\bfseries Supplemental Material for\\
``Localization in microcavities revealed by phase-space non-Hermitian skin effect''\par}
\vspace{1.2em}
\endgroup


\section{Additional resonance modes within narrow frequency windows}
\label{sec:additional_modes}

This section presents additional resonance modes of two-dimensional spiral dielectric microcavities together with the corresponding eigenstates of the closed spiral billiard within narrow frequency windows around \(nkR\simeq100\). The dielectric cavity is surrounded by air and consists of a spiral arc connected by a straight notch. For transverse-magnetic (TM) polarization, the resonance modes are obtained by solving the two-dimensional Helmholtz equation,
\begin{equation}
\left[\nabla^2+n^2(\mathbf r)k^2\right]\psi(\mathbf r)=0,
\end{equation}
where \(n(\mathbf r)\) is the refractive-index profile. Purely outgoing-wave boundary conditions are imposed outside the cavity, yielding complex resonance frequencies \(nkR\).

The resonance modes are described by the electric field
\(\psi(\mathbf r)=E_z(\mathbf r)\), whose real-space amplitude patterns are given by \(|\psi(\mathbf r)|\). Their localization in dielectric phase space is visualized by the corresponding boundary Husimi distributions \(H(s,p)\), where \(s\) is the boundary arclength and \(p=\sin\chi\) is the Birkhoff momentum coordinate.

Figures~S1-S3 show forty resonance amplitude patterns for spiral dielectric microcavities with refractive indices \(n=2\), \(3\), and \(4\), respectively. Figure~S4 shows forty eigenstate amplitude patterns of the corresponding closed spiral billiard within the same frequency range. The corresponding boundary Husimi distributions are presented in Figs.~S5-S7 for the dielectric resonances and in Fig.~S8 for the closed-billiard eigenstates. These additional examples demonstrate that localization near the critical line is a common feature of the dielectric resonances, whereas the corresponding closed billiard exhibits only weak localization without comparable systematic accumulation near the critical line.

\begin{figure*}[t]
\centering
\includegraphics[width=\linewidth]{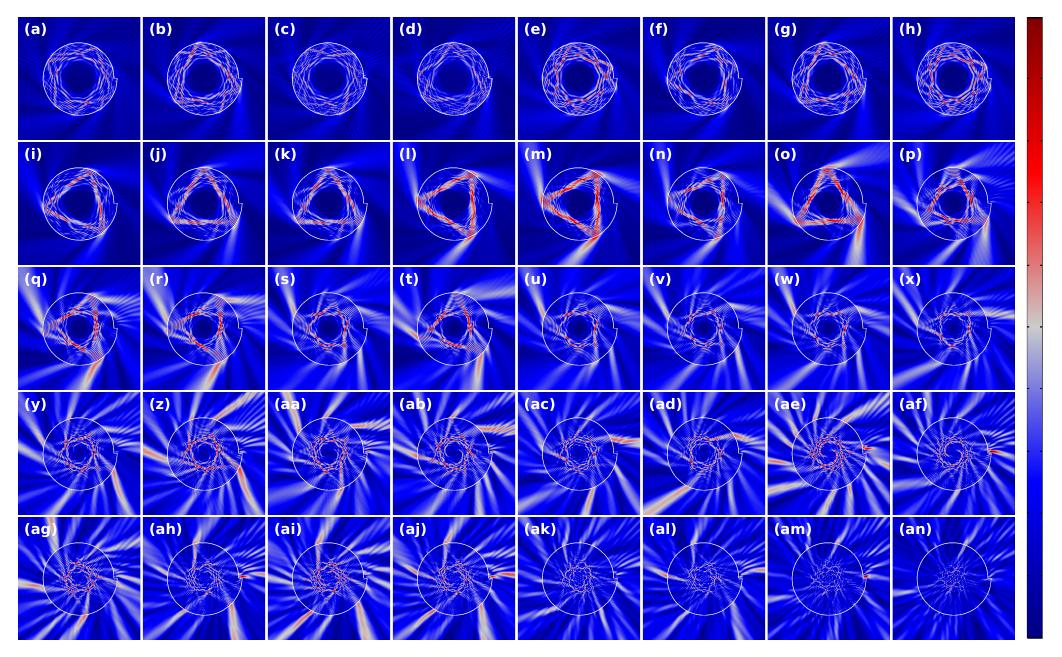}
\caption{
Resonance amplitude patterns within a narrow frequency window around \(nkR\simeq100\) for the spiral dielectric microcavity with refractive index \(n=2\).
The corresponding complex resonance frequencies \(nkR\) are
(a) \(99.7287-0.0432i\),
(b) \(99.9721-0.0441i\),
(c) \(100.1539-0.0447i\),
(d) \(100.1469-0.0465i\),
(e) \(99.8204-0.0528i\),
(f) \(99.7220-0.0538i\),
(g) \(99.9901-0.0544i\),
(h) \(99.8139-0.0587i\),
(i) \(99.6440-0.0656i\),
(j) \(100.2991-0.0983i\),
(k) \(100.2800-0.1082i\),
(l) \(99.9456-0.1196i\),
(m) \(99.9492-0.1442i\),
(n) \(100.0638-0.1589i\),
(o) \(100.1828-0.1847i\),
(p) \(100.1587-0.2338i\),
(q) \(99.6970-0.2635i\),
(r) \(99.7360-0.2933i\),
(s) \(100.1174-0.2981i\),
(t) \(100.3366-0.2984i\),
(u) \(100.0267-0.3112i\),
(v) \(99.9527-0.3465i\),
(w) \(100.1244-0.3542i\),
(x) \(99.6558-0.3716i\),
(y) \(99.8719-0.3870i\),
(z) \(99.7959-0.3888i\),
(aa) \(99.7856-0.3893i\),
(ab) \(99.9007-0.3959i\),
(ac) \(99.7094-0.4003i\),
(ad) \(100.2618-0.4015i\),
(ae) \(100.0136-0.4299i\),
(af) \(100.1138-0.4313i\),
(ag) \(100.0729-0.4343i\),
(ah) \(100.2881-0.4403i\),
(ai) \(100.1901-0.4461i\),
(aj) \(100.3106-0.4484i\),
(ak) \(99.9947-0.4574i\),
(al) \(100.1738-0.4591i\),
(am) \(99.7074-0.4634i\), and
(an) \(99.8655-0.4865i\).
The resonances are arranged in descending order of \(\operatorname{Im}(nkR)\). The color scale represents the normalized amplitude \(|\psi|\), with \(\max|\psi|=1\) for each state.
}
\label{fig:supp_cavity_n2}
\end{figure*}

\begin{figure*}[t]
\centering
\includegraphics[width=\linewidth]{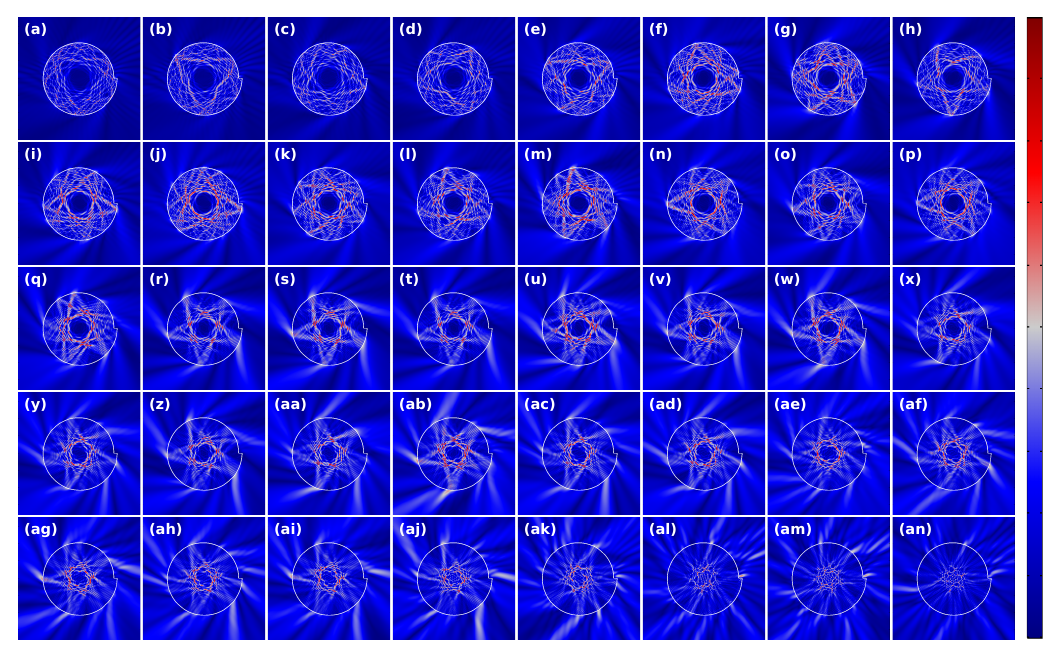}
\caption{
Resonance amplitude patterns within a narrow frequency window around \(nkR\simeq100\) for the spiral dielectric microcavity with refractive index \(n=3\).
The corresponding complex resonance frequencies \(nkR\) are
(a) \(99.9089-0.0180i\),
(b) \(99.8876-0.0198i\),
(c) \(100.1006-0.0221i\),
(d) \(100.1275-0.0225i\),
(e) \(99.7463-0.0347i\),
(f) \(100.2142-0.0376i\),
(g) \(99.6583-0.0389i\),
(h) \(99.7517-0.0396i\),
(i) \(99.9725-0.0415i\),
(j) \(99.9882-0.0468i\),
(k) \(99.8034-0.0490i\),
(l) \(100.2469-0.0501i\),
(m) \(100.2812-0.0545i\),
(n) \(100.0391-0.0605i\),
(o) \(99.8341-0.0650i\),
(p) \(100.0765-0.0709i\),
(q) \(100.3284-0.0778i\),
(r) \(99.7460-0.1052i\),
(s) \(99.8008-0.1072i\),
(t) \(99.6805-0.1142i\),
(u) \(100.3212-0.1161i\),
(v) \(99.8688-0.1221i\),
(w) \(99.7869-0.1253i\),
(x) \(100.2659-0.1320i\),
(y) \(100.1838-0.1470i\),
(z) \(99.9440-0.1497i\),
(aa) \(100.0774-0.1534i\),
(ab) \(99.6929-0.1639i\),
(ac) \(100.1525-0.1677i\),
(ad) \(100.2107-0.1678i\),
(ae) \(99.8520-0.1796i\),
(af) \(100.2915-0.1856i\),
(ag) \(99.9797-0.1964i\),
(ah) \(100.0509-0.2085i\),
(ai) \(100.0949-0.2094i\),
(aj) \(100.1846-0.2339i\),
(ak) \(99.9943-0.2370i\),
(al) \(99.6551-0.2578i\),
(am) \(100.3442-0.2630i\), and
(an) \(99.8299-0.2740i\).
The resonances are arranged in descending order of \(\operatorname{Im}(nkR)\). The color scale represents the normalized amplitude \(|\psi|\), with \(\max|\psi|=1\) for each state.
}
\label{fig:supp_cavity_n3}
\end{figure*}

\begin{figure*}[t]
\centering
\includegraphics[width=\linewidth]{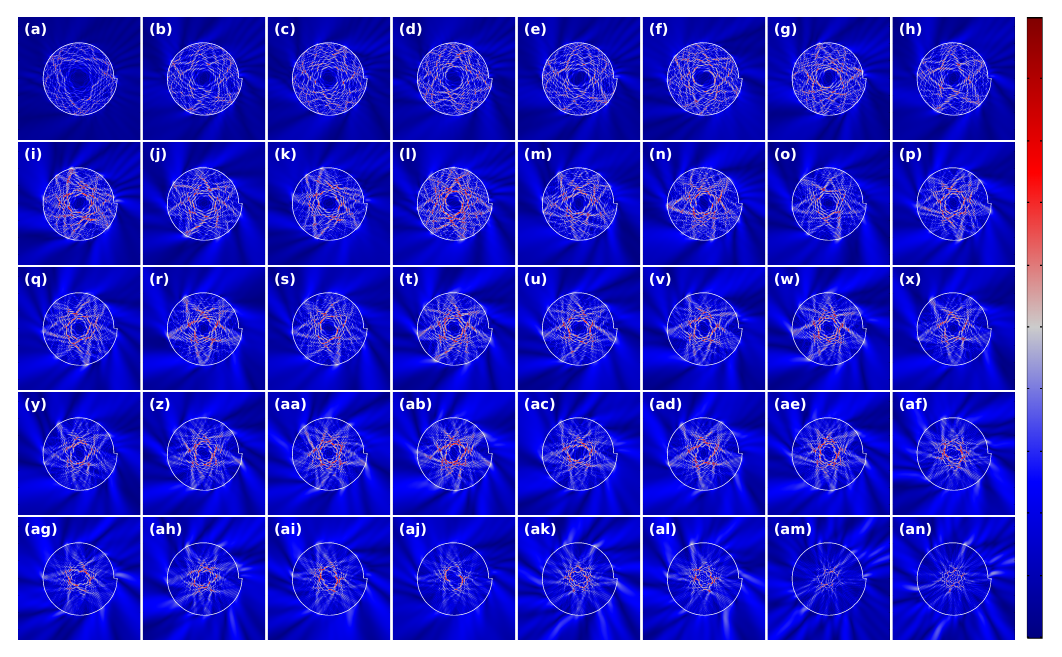}
\caption{
Resonance amplitude patterns within a narrow frequency window around \(nkR\simeq100\) for the spiral dielectric microcavity with refractive index \(n=4\).
The corresponding complex resonance frequencies \(nkR\) are
(a) \(99.9563-0.0092i\),
(b) \(99.6880-0.0191i\),
(c) \(100.1945-0.0193i\),
(d) \(100.1634-0.0207i\),
(e) \(99.7144-0.0215i\),
(f) \(99.9251-0.0237i\),
(g) \(100.2843-0.0256i\),
(h) \(99.7892-0.0286i\),
(i) \(100.3225-0.0316i\),
(j) \(99.8447-0.0345i\),
(k) \(100.3467-0.0345i\),
(l) \(100.0504-0.0355i\),
(m) \(99.7706-0.0372i\),
(n) \(100.0645-0.0410i\),
(o) \(100.0025-0.0458i\),
(p) \(100.1505-0.0493i\),
(q) \(100.1149-0.0496i\),
(r) \(99.7996-0.0498i\),
(s) \(100.0806-0.0505i\),
(t) \(99.8824-0.0536i\),
(u) \(99.8274-0.0553i\),
(v) \(99.9717-0.0598i\),
(w) \(99.9557-0.0634i\),
(x) \(99.6698-0.0636i\),
(y) \(99.7304-0.0644i\),
(z) \(100.2275-0.0689i\),
(aa) \(100.2840-0.0721i\),
(ab) \(100.3157-0.0733i\),
(ac) \(99.8605-0.0749i\),
(ad) \(99.6471-0.0753i\),
(ae) \(100.2223-0.0821i\),
(af) \(99.7368-0.1012i\),
(ag) \(100.1114-0.1042i\),
(ah) \(99.9413-0.1051i\),
(ai) \(100.1877-0.1052i\),
(aj) \(100.0488-0.1122i\),
(ak) \(100.0080-0.1351i\),
(al) \(100.1670-0.1359i\),
(am) \(100.3352-0.1680i\), and
(an) \(99.8278-0.1698i\).
The resonances are arranged in descending order of \(\operatorname{Im}(nkR)\). The color scale represents the normalized amplitude \(|\psi|\), with \(\max|\psi|=1\) for each state.
}
\label{fig:supp_cavity_n4}
\end{figure*}

\begin{figure*}[t]
\centering
\includegraphics[width=\linewidth]{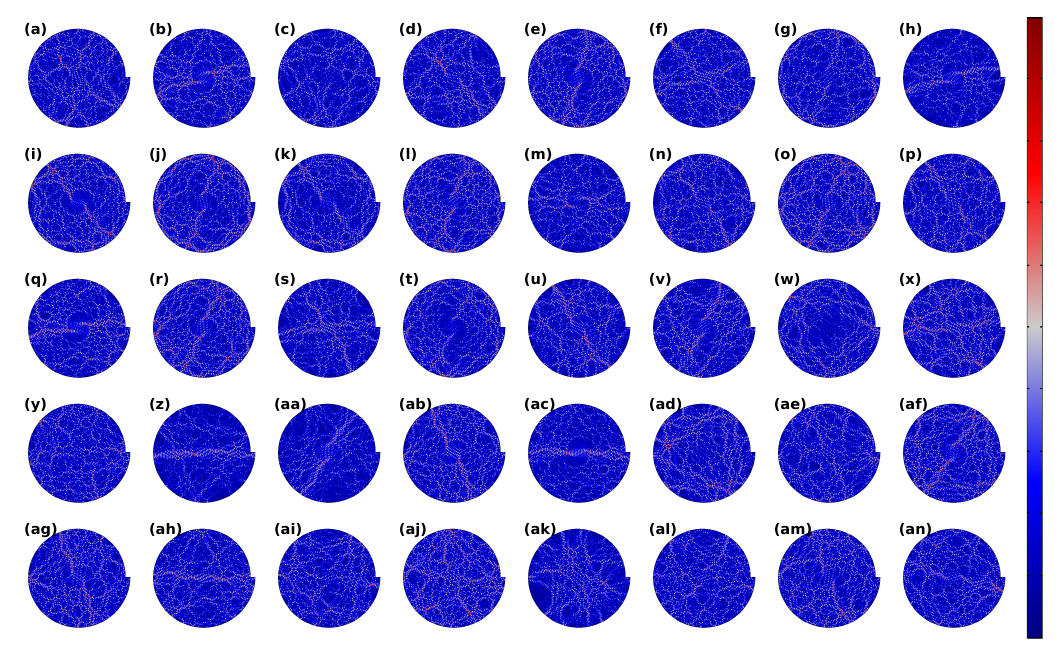}
\caption{
Eigenstate amplitude patterns within the corresponding frequency window around \(nkR\simeq100\) for the closed spiral billiard.
The corresponding eigenfrequencies \(nkR\) are
(a) \(99.6426\),
(b) \(99.6638\),
(c) \(99.6744\),
(d) \(99.7030\),
(e) \(99.7190\),
(f) \(99.7351\),
(g) \(99.7623\),
(h) \(99.7757\),
(i) \(99.7793\),
(j) \(99.8054\),
(k) \(99.8290\),
(l) \(99.8514\),
(m) \(99.8582\),
(n) \(99.8776\),
(o) \(99.8995\),
(p) \(99.9096\),
(q) \(99.9408\),
(r) \(99.9486\),
(s) \(99.9807\),
(t) \(99.9960\),
(u) \(100.0031\),
(v) \(100.0309\),
(w) \(100.0481\),
(x) \(100.0575\),
(y) \(100.0839\),
(z) \(100.1074\),
(aa) \(100.1106\),
(ab) \(100.1263\),
(ac) \(100.1581\),
(ad) \(100.1734\),
(ae) \(100.1809\),
(af) \(100.2092\),
(ag) \(100.2241\),
(ah) \(100.2495\),
(ai) \(100.2593\),
(aj) \(100.2828\),
(ak) \(100.2971\),
(al) \(100.3124\),
(am) \(100.3316\), and
(an) \(100.3621\).
The eigenstates are arranged in ascending order of \(nkR\). The color scale represents the normalized amplitude \(|\psi|\), with \(\max|\psi|=1\) for each state.
}
\label{fig:supp_billiard_n2}
\end{figure*}

\begin{figure*}[t]
\centering
\includegraphics[width=\linewidth]{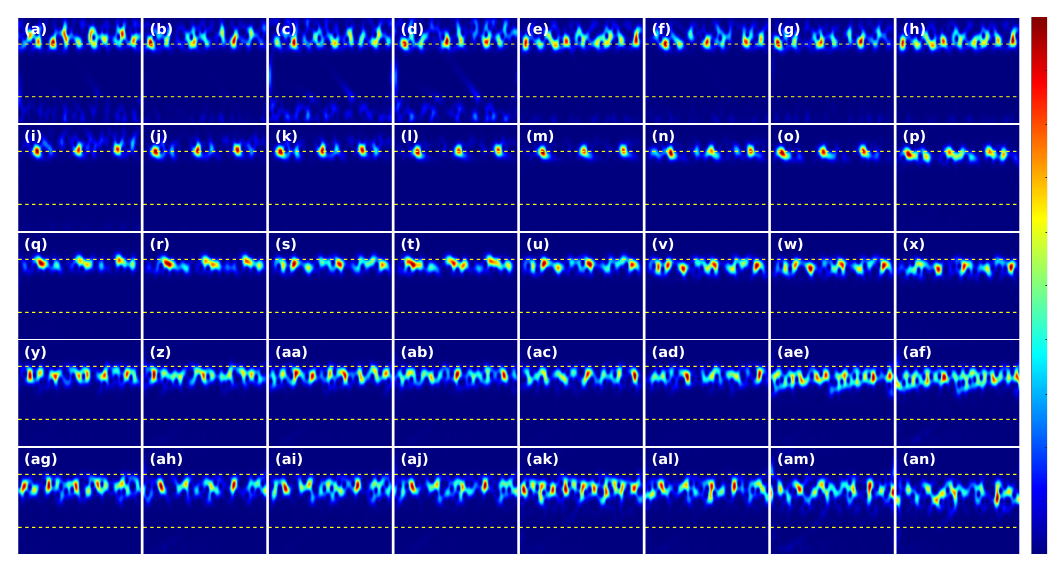}
\caption{
Boundary Husimi distributions of the resonances shown in Fig.~\ref{fig:supp_cavity_n2} for the spiral dielectric microcavity with \(n=2\).
Panels (a)-(an) correspond to the amplitude patterns in Fig.~\ref{fig:supp_cavity_n2} and are arranged in the same order.
In each panel, the horizontal coordinate is the boundary arclength \(s\), and the vertical coordinate is the Birkhoff momentum coordinate \(p=\sin\chi\).
The dashed horizontal lines indicate the critical lines for total internal reflection, \(p=p_c= \pm 1/2\).
The concentration of Husimi intensity near the critical line directly visualizes the systematic critical-line localization discussed in the main text. The color scale represents the normalized Husimi distribution \(H(s,p)\), with \(\max H(s,p)=1\) for each state.
}
\label{fig:supp_cavity_husimi_n2}
\end{figure*}

\begin{figure*}[t]
\centering
\includegraphics[width=\linewidth]{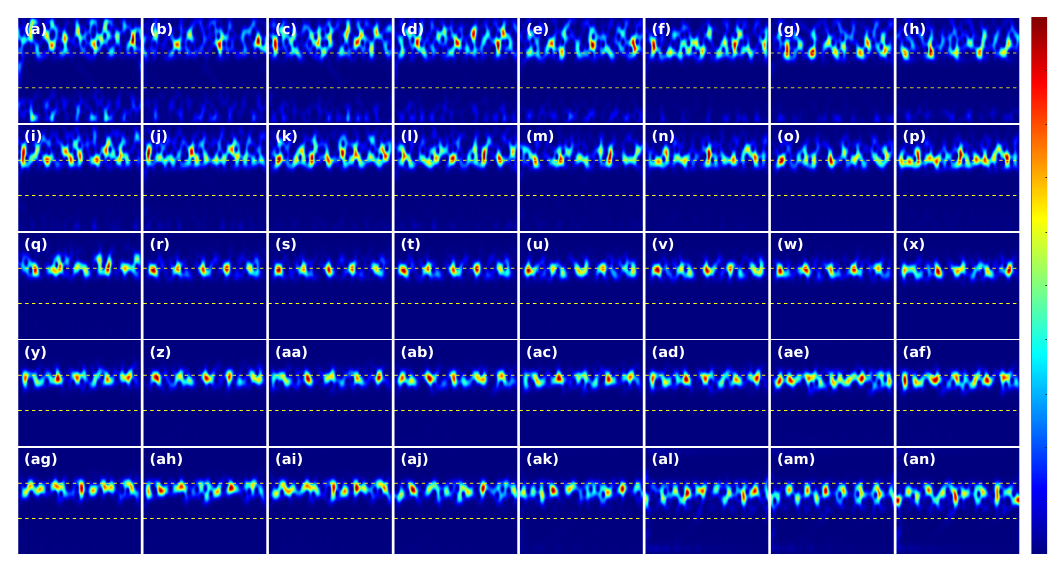}
\caption{
Boundary Husimi distributions of the resonances shown in Fig.~\ref{fig:supp_cavity_n3} for the spiral dielectric microcavity with \(n=3\).
Panels (a)-(an) correspond to the amplitude patterns in Fig.~\ref{fig:supp_cavity_n3} and are arranged in the same order.
In each panel, the horizontal coordinate is the boundary arclength \(s\), and the vertical coordinate is the Birkhoff momentum coordinate \(p=\sin\chi\).
The dashed horizontal lines indicate the critical lines for total internal reflection, \(p=p_c= \pm 1/3\).
The concentration of Husimi intensity near the critical line directly visualizes the systematic critical-line localization discussed in the main text. The color scale represents the normalized Husimi distribution \(H(s,p)\), with \(\max H(s,p)=1\) for each state.
}
\label{fig:supp_cavity_husimi_n3}
\end{figure*}

\begin{figure*}[t]
\centering
\includegraphics[width=\linewidth]{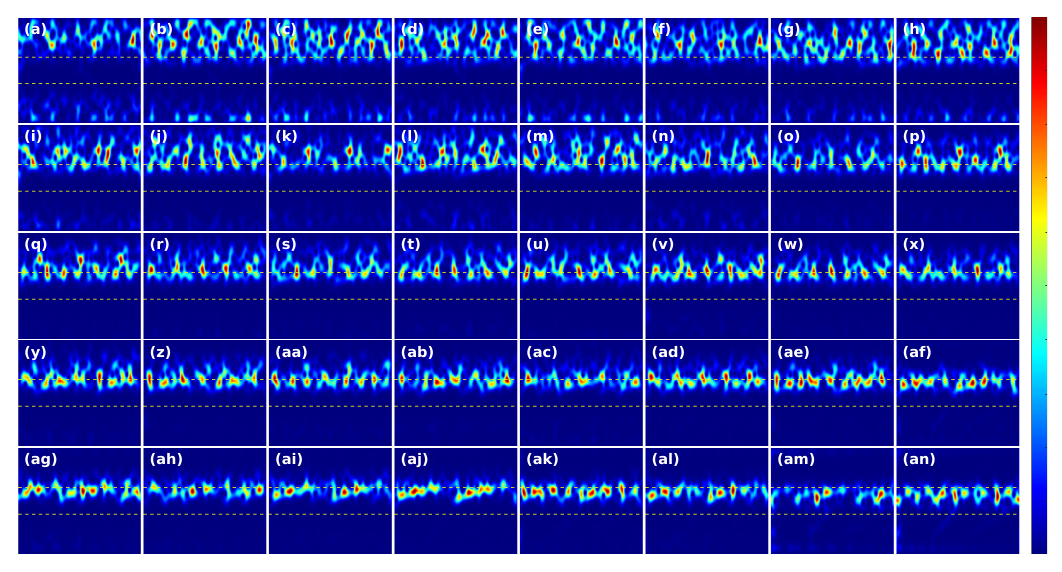}
\caption{
Boundary Husimi distributions of the resonances shown in Fig.~\ref{fig:supp_cavity_n4} for the spiral dielectric microcavity with \(n=4\).
Panels (a)-(an) correspond to the amplitude patterns in Fig.~\ref{fig:supp_cavity_n4} and are arranged in the same order.
In each panel, the horizontal coordinate is the boundary arclength \(s\), and the vertical coordinate is the Birkhoff momentum coordinate \(p=\sin\chi\).
The dashed horizontal lines indicate the critical lines for total internal reflection, \(p=p_c= \pm 1/4\).
The concentration of Husimi intensity near the critical line directly visualizes the systematic critical-line localization discussed in the main text. The color scale represents the normalized Husimi distribution \(H(s,p)\), with \(\max H(s,p)=1\) for each state.
}
\label{fig:supp_cavity_husimi_n4}
\end{figure*}

\begin{figure*}[t]
\centering
\includegraphics[width=\linewidth]{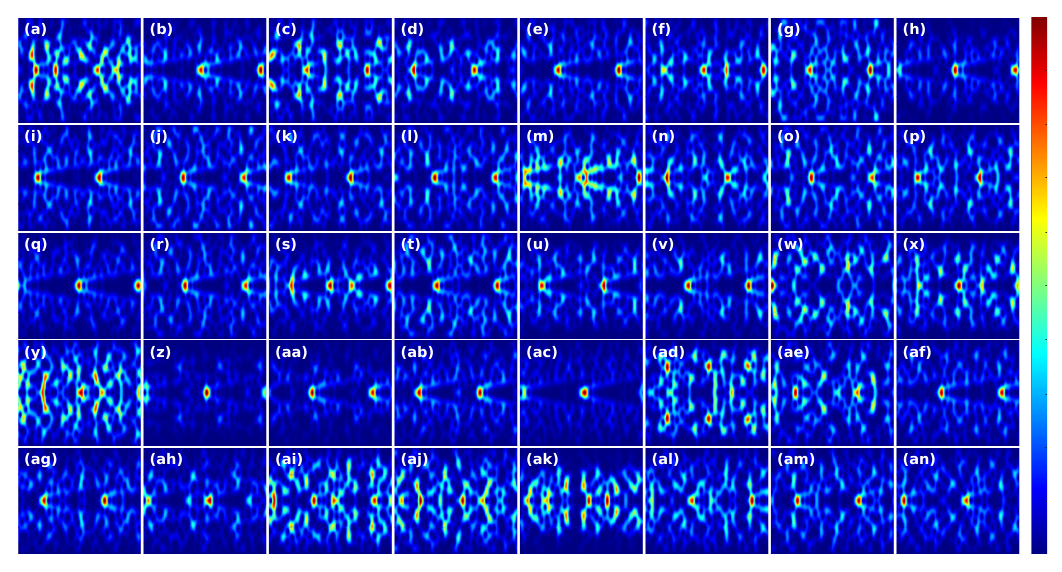}
\caption{
Boundary Husimi distributions of the eigenstates shown in Fig.~\ref{fig:supp_billiard_n2} for the corresponding closed spiral billiard.
Panels (a)-(an) correspond to the amplitude patterns in Fig.~\ref{fig:supp_billiard_n2} and are arranged in the same order.
In each panel, the horizontal coordinate is the boundary arclength \(s\), and the vertical coordinate is the Birkhoff momentum coordinate \(p=\sin\chi\).
Unlike the dielectric resonances, the closed-billiard eigenstates do not exhibit systematic accumulation near a particular momentum line. The color scale represents the normalized Husimi distribution \(H(s,p)\), with \(\max H(s,p)=1\) for each state.
}
\label{fig:supp_billiard_husimi_n2}
\end{figure*}

\clearpage

\section{Detailed evaluation of the effective hopping matrix elements}
\label{sec:detailed_hopping}

The derivation summarized in End Matter B shows that directed wave propagation
with finite spatial extent generates asymmetric local couplings when projected
onto localized basis states.
Here, we provide the detailed evaluation of the projected matrix elements used
in that derivation.

The propagation kernel is taken to be spatially localized with a finite width,
\begin{equation}
K(x,x')
=
G(x-x'-a),
\label{eq:supp_wave_kernel}
\end{equation}
where
\begin{equation}
G(y)
=
\frac{1}{\sqrt{2\pi}w}
\exp
\left(
-\frac{y^2}{2w^2}
\right),
\label{eq:supp_gaussian_kernel}
\end{equation}
\(w\) is the width parameter of the propagation kernel, and \(a>0\) is the
displacement of its center along the directed propagation.
We introduce normalized Gaussian basis states centered at \(x_j=jd\),
\begin{equation}
\phi_j(x)
=
\frac{1}{(\pi\sigma^2)^{1/4}}
\exp
\left[
-\frac{(x-x_j)^2}{2\sigma^2}
\right],
\qquad
x_j=jd,
\label{eq:supp_gaussian_basis}
\end{equation}
where \(d\) is the spacing between neighboring basis-state centers and
\(\sigma\) is the width parameter of each localized basis state.

The projected matrix elements of the one-step propagation operator are
\begin{align}
t_{ij}
&=
\langle\phi_i|K|\phi_j\rangle
\nonumber\\
&=
\int dx\,dx'\,
\phi_i^*(x)
G(x-x'-a)
\phi_j(x').
\label{eq:supp_effective_hopping}
\end{align}
Substituting Eqs.~(\ref{eq:supp_gaussian_kernel}) and
(\ref{eq:supp_gaussian_basis}) gives
\begin{align}
t_{ij}
&=
\frac{1}{\sqrt{\pi}\sigma}
\frac{1}{\sqrt{2\pi}w}
\int dx\,dx'
\exp
\left[
-\frac{(x-x_i)^2}{2\sigma^2}
-\frac{(x'-x_j)^2}{2\sigma^2}
\right.
\left.
-\frac{(x-x'-a)^2}{2w^2}
\right].
\label{eq:supp_hopping_integral}
\end{align}

We introduce the shifted variables
\begin{equation}
u=x-x_i,
\qquad
v=x'-x_j,
\qquad
\Delta_{ij}=x_i-x_j-a.
\label{eq:supp_shifted_variables}
\end{equation}
The matrix element then becomes
\begin{align}
t_{ij}
&=
\frac{1}{\sqrt{\pi}\sigma}
\frac{1}{\sqrt{2\pi}w}
\int du\,dv\,
\exp
\left[
-\frac{u^2}{2\sigma^2}
-\frac{v^2}{2\sigma^2}
-\frac{(u-v+\Delta_{ij})^2}{2w^2}
\right].
\label{eq:supp_shifted_integral}
\end{align}

To separate the two integration variables, we define
\begin{equation}
q=u-v,
\qquad
r=\frac{u+v}{2}.
\label{eq:supp_relative_coordinates}
\end{equation}
The inverse transformation is
\begin{equation}
u=r+\frac{q}{2},
\qquad
v=r-\frac{q}{2},
\end{equation}
and its Jacobian has unit magnitude.
Moreover,
\begin{equation}
u^2+v^2
=
2r^2+\frac{q^2}{2}.
\label{eq:supp_coordinate_identity}
\end{equation}
Using these relations, Eq.~(\ref{eq:supp_shifted_integral}) becomes
\begin{align}
t_{ij}
&=
\frac{1}{\sqrt{\pi}\sigma}
\frac{1}{\sqrt{2\pi}w}
\int dr\,
\exp
\left(
-\frac{r^2}{\sigma^2}
\right)
\times
\int dq\,
\exp
\left[
-\frac{q^2}{4\sigma^2}
-\frac{(q+\Delta_{ij})^2}{2w^2}
\right].
\label{eq:supp_separated_integral}
\end{align}
The integration over \(r\) gives
\begin{equation}
\int_{-\infty}^{\infty}
dr\,
\exp
\left(
-\frac{r^2}{\sigma^2}
\right)
=
\sqrt{\pi}\sigma.
\label{eq:supp_r_integral}
\end{equation}

To evaluate the remaining integral, we define the effective width parameter
\begin{equation}
\Sigma^2=w^2+2\sigma^2.
\label{eq:supp_effective_width}
\end{equation}
The exponent involving \(q\) can then be written as
\begin{align}
-\frac{q^2}{4\sigma^2}
-\frac{(q+\Delta_{ij})^2}{2w^2}
&=
-\frac{\Sigma^2}{4\sigma^2w^2}
\left(
q+
\frac{2\sigma^2\Delta_{ij}}{\Sigma^2}
\right)^2
-\frac{\Delta_{ij}^2}{2\Sigma^2}.
\label{eq:supp_complete_square}
\end{align}
The \(q\) integral is therefore
\begin{align}
&\int_{-\infty}^{\infty}
dq\,
\exp
\left[
-\frac{q^2}{4\sigma^2}
-\frac{(q+\Delta_{ij})^2}{2w^2}
\right]
\nonumber\\
&\qquad=
\exp
\left(
-\frac{\Delta_{ij}^2}{2\Sigma^2}
\right)
\int_{-\infty}^{\infty}
dq\,
\exp
\left[
-\frac{\Sigma^2}{4\sigma^2w^2}
\left(
q+
\frac{2\sigma^2\Delta_{ij}}{\Sigma^2}
\right)^2
\right]
\nonumber\\
&\qquad=
\frac{2\sqrt{\pi}\sigma w}{\Sigma}
\exp
\left(
-\frac{\Delta_{ij}^2}{2\Sigma^2}
\right).
\label{eq:supp_q_integral}
\end{align}

Combining Eqs.~(\ref{eq:supp_separated_integral}),
(\ref{eq:supp_r_integral}), and
(\ref{eq:supp_q_integral}) yields
\begin{equation}
t_{ij}
=
t_0
\exp
\left[
-\frac{(x_i-x_j-a)^2}{2\Sigma^2}
\right],
\label{eq:supp_gaussian_hopping}
\end{equation}
where
\begin{equation}
t_0
=
\frac{\sqrt{2}\sigma}{\Sigma}.
\label{eq:supp_hopping_prefactor}
\end{equation}
Equation~(\ref{eq:supp_gaussian_hopping}) shows that the coupling is largest
when the displacement between the basis-state centers matches the displacement
\(a\) of the propagation-kernel center.

For neighboring basis states, the coupling along the directed propagation is
obtained from \(x_i-x_j=d\),
\begin{equation}
t_R
=
t_0
\exp
\left[
-\frac{(d-a)^2}{2\Sigma^2}
\right],
\label{eq:supp_right_hopping}
\end{equation}
whereas the coupling in the opposite direction follows from
\(x_i-x_j=-d\),
\begin{equation}
t_L
=
t_0
\exp
\left[
-\frac{(d+a)^2}{2\Sigma^2}
\right].
\label{eq:supp_left_hopping}
\end{equation}
Their ratio is
\begin{equation}
\frac{t_R}{t_L}
=
\exp
\left(
\frac{2ad}{\Sigma^2}
\right).
\label{eq:supp_hopping_ratio}
\end{equation}
Thus, \(a>0\) gives \(t_R>t_L\), whereas \(a=0\) gives \(t_R=t_L\).
The hopping asymmetry therefore originates directly from the directed
displacement of the propagation kernel relative to the localized basis states.

The nearest-neighbor couplings may equivalently be written in the standard
Hatano-Nelson form,
\begin{equation}
t_R=t e^{g_{\mathrm{eff}}},
\qquad
t_L=t e^{-g_{\mathrm{eff}}},
\end{equation}
with
\begin{equation}
t
=
t_0
\exp
\left[
-\frac{d^2+a^2}{2\Sigma^2}
\right],
\qquad
g_{\mathrm{eff}}
=
\frac{ad}{\Sigma^2}.
\end{equation}
This completes the detailed evaluation of the effective hopping matrix
elements used in End Matter B and the main text.

\section{Determination of the localization momentum \(p_{\max}\)}
\label{sec:pmax}

To characterize the dominant localization position in dielectric phase space, the boundary Husimi distribution \(H(s,p)\) is averaged over the boundary coordinate \(s\),
\begin{equation}
\overline{H}(p)
=
\frac{1}{L_{\partial}}
\int_{0}^{L_{\partial}}
H(s,p)\,ds,
\label{eq:supp_H_average}
\end{equation}
where \(L_{\partial}\) is the total boundary length.
The momentum coordinate at which the averaged Husimi distribution reaches its maximum is defined as
\begin{equation}
p_{\max}
=
\operatorname*{arg\,max}_{p}
\overline{H}(p),
\label{eq:supp_pmax}
\end{equation}
whereas the averaged momentum is
\begin{equation}
\langle p\rangle
=
\frac{
\displaystyle\int p\,\overline{H}(p)\,dp
}{
\displaystyle\int \overline{H}(p)\,dp
}.
\label{eq:supp_paverage}
\end{equation}
The quantity \(p_{\max}\) is used in the statistical analysis presented in Fig.~4 of the main text because it directly identifies the momentum coordinate of the dominant localization.

\begin{figure}[t]
\centering
\includegraphics[width=0.95\linewidth]{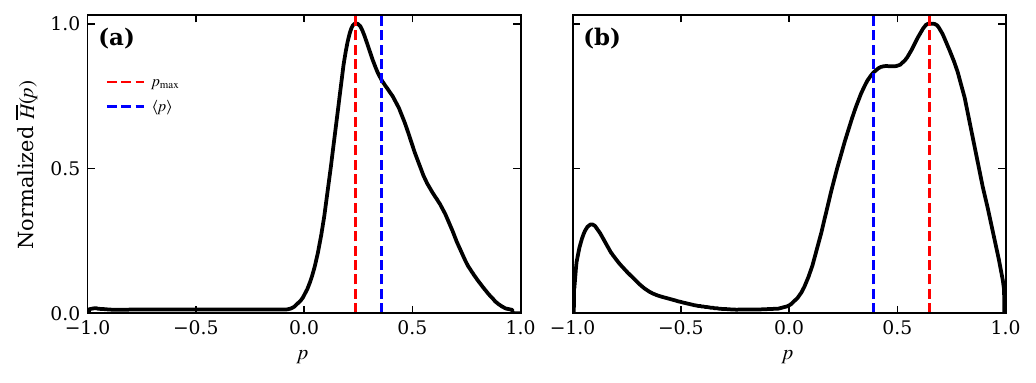}
\caption{
Representative averaged boundary Husimi distributions,
\(\overline{H}(p)\), for two resonances of the spiral dielectric microcavity with \(n=4\).
The corresponding two-dimensional boundary Husimi distributions are shown in Figs.~S7(q) and S7(d), respectively.
The red dashed line indicates \(p_{\max}\), while the blue dashed line indicates the averaged momentum \(\langle p\rangle\).
(a) An asymmetric localization profile shifts the averaged momentum away from the dominant localization.
(b) An additional weaker peak at negative momentum further shifts the averaged momentum away from the dominant localization.
These examples illustrate why \(p_{\max}\) provides a more reliable characterization of the dominant localization than the averaged momentum.
}
\label{fig:pmax_examples}
\end{figure}

Figure~\ref{fig:pmax_examples} illustrates two representative examples corresponding to the boundary Husimi distributions shown in Figs.~S7(q) and S7(d). Resonances localized near the critical line generally exhibit asymmetric localization profiles extending differently into the total-internal-reflection and refractive regions. In Fig.~\ref{fig:pmax_examples}(a), this asymmetry shifts the averaged momentum toward larger positive momentum, although the localization remains centered at \(p_{\max}\). In Fig.~\ref{fig:pmax_examples}(b), a weaker secondary peak at negative momentum shifts the averaged momentum even further away from the dominant localization. In both cases, \(p_{\max}\) correctly identifies the dominant localization, whereas the averaged momentum is influenced by the overall shape of the distribution.

These examples demonstrate that the averaged momentum may obscure the dominant localization either because of the asymmetric localization profile near the critical line or because of additional secondary peaks. Since the resonances included in the statistical analysis are typically characterized by a single dominant Husimi peak near the positive critical line, \(p_{\max}\) provides a direct and robust measure of the localization position relevant to the generalized non-Hermitian skin effect.

\section{Hatano-Nelson model with a dielectric Fresnel-loss profile}
\label{sec:fresnel_loss}

The inhomogeneous-loss Hatano-Nelson model discussed in the main text employs a uniform onsite loss in the lossy region in order to isolate the essential mechanism responsible for interface localization.
In an actual dielectric microcavity, however, refractive escape varies continuously with the angle of incidence according to the Fresnel reflection coefficient.
Here, we examine how this continuous dielectric loss profile modifies the localization properties.

For TM polarization, the amplitude reflection coefficient for incidence from a dielectric medium with refractive index n into air is given by
\begin{equation}
r_{\mathrm{TM}}(p)
=
\frac{
n\sqrt{1-p^2}
-
\sqrt{1-n^2p^2}
}{
n\sqrt{1-p^2}
+
\sqrt{1-n^2p^2}
},
\qquad
|p|<\frac{1}{n}.
\label{eq:supp_fresnel_coefficient}
\end{equation}
For \(|p|\geq 1/n\), total internal reflection occurs and
\(\left|r_{\mathrm{TM}}(p)\right|=1\).
The corresponding amplitude-loss parameter is defined as
\begin{equation}
\gamma_{\mathrm F}(p)
=
-\ln\left|r_{\mathrm{TM}}(p)\right|
=
-\frac{1}{2}\ln R_{\mathrm{TM}}(p),
\label{eq:supp_fresnel_loss}
\end{equation}
where
\begin{equation}
R_{\mathrm{TM}}(p)
=
\left|r_{\mathrm{TM}}(p)\right|^2
\end{equation}
is the Fresnel reflectance.
Accordingly, \(\gamma_{\mathrm F}(p)=0\) in the total-internal-reflection region \(|p|\geq 1/n\), while it varies continuously with \(p\) in the refractive region.

\begin{figure}[t]
\centering
\includegraphics[width=\linewidth]{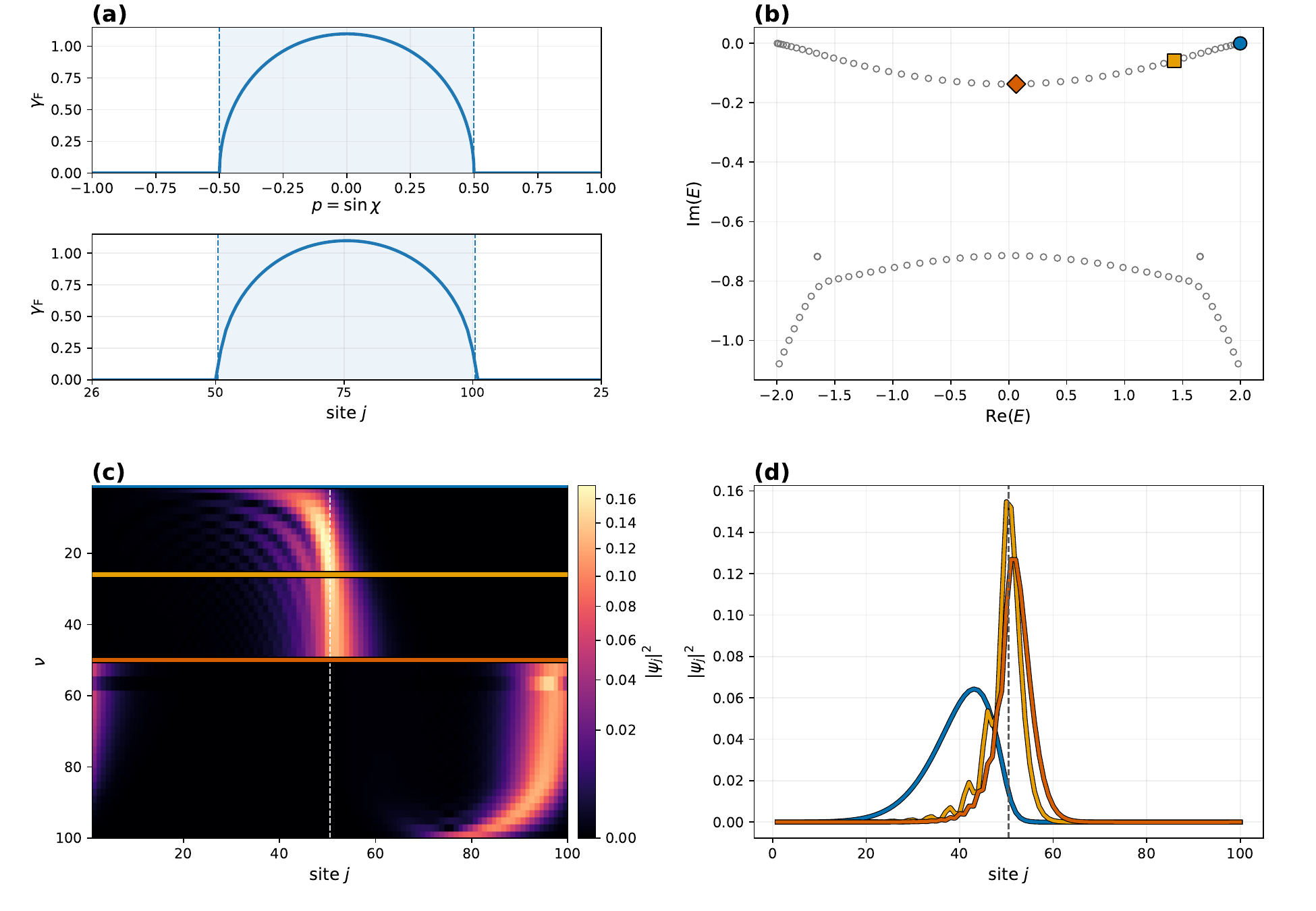}
\caption{
Hatano-Nelson model with a dielectric TM Fresnel-loss profile.
(a) Momentum-dependent amplitude-loss profile
\(\gamma_{\mathrm F}(p)\)
obtained from the TM Fresnel reflection coefficient and the corresponding site-dependent loss profile
\(\gamma_{\mathrm F}(j)\)
used in the lattice Hamiltonian.
The shaded region denotes the lossy half of the lattice.
(b) Complex-energy spectrum.
(c) Spatial intensities of all right eigenstates, indexed by \(\nu\) in order of decreasing
\(\operatorname{Im}(E)\).
(d) Representative right eigenstates selected from the weakly decaying branch.
Interface localization remains robust under the realistic dielectric loss profile.
Compared with the uniform-loss model, the localization centers of the eigenstates with larger decay rates within the weakly decaying branch shift slightly away from the interface and into the lossy region.
}
\label{fig:supp_fresnel_loss_HN}
\end{figure}

The momentum interval
\begin{equation}
-\frac{1}{n}
\leq p
\leq
\frac{1}{n},
\end{equation}
corresponding to the refractive region of the dielectric cavity, is mapped uniformly onto the lossy half of the Hatano-Nelson ring.
The uniform onsite loss used in the main text is then replaced by the site-dependent loss profile
\begin{equation}
-i\gamma
\longrightarrow
-i\gamma_{\mathrm F}(j).
\label{eq:supp_site_loss}
\end{equation}
Figure~\ref{fig:supp_fresnel_loss_HN} summarizes the corresponding loss profile, complex-energy spectrum, and right eigenstates.

Although the sharp loss interface of the uniform-loss model is replaced by a continuous dielectric loss profile, Figs.~\ref{fig:supp_fresnel_loss_HN}(b)-\ref{fig:supp_fresnel_loss_HN}(d) show that interface localization remains robust.
The complex-energy spectrum continues to separate into weakly and strongly decaying branches, and the long-lived eigenstates remain accumulated near the interface between the lossless and lossy regions.

Compared with the uniform-loss model, the localization centers of the eigenstates with larger decay rates within the weakly decaying branch shift slightly away from the interface and into the lossy region.
This shift originates from the continuous variation of the dielectric loss across the refractive region rather than from an abrupt loss boundary.
Therefore, although the generalized non-Hermitian skin effect continues to determine the localization mechanism, the localization maximum is not required to occur exactly at the effective interface.
This behavior directly accounts for the resonance statistics of spiral dielectric microcavities: long-lived resonances predominantly accumulate near the critical line, whereas relatively shorter-lived resonances within the selected set may exhibit \(p_{\max}<p_c\).

\end{document}